\documentstyle[12pt,aps,epsfig]{revtex}

\newcommand{\be}{\begin{equation}}
\newcommand{\ee}{\end{equation}}
\newcommand{\bea}{\begin{eqnarray}}
\newcommand{\eea}{\end{eqnarray}}
\newcommand{\gv}{\vec{\gamma}}
\newcommand{\pv}{{\bf p}}
\newcommand{\kv}{{\bf k}}
\newcommand{\qv}{{\bf q}}
\newcommand{\pvuni}{\hat{{\bf p}}}

\newcommand{\kvuni}{\hat{{\bf k}}}
\newcommand{\pkvuni}{\widehat{{\bf p}+{\bf k}}}
\newcommand{\puni}{\hat{p}}

\newcommand{\kuni}{\hat{k}}
\newcommand{\pkuni}{\widehat{p+k}}
\newcommand{\modp}{\vert {\bf p} \vert}
\newcommand{\modk}{\vert {\bf k} \vert}

\newcommand{\sg}[1]{\mbox{sgn$(#1)$}}

\newcommand{\cP}{{\cal P}}
\newcommand{\cD}{{\cal D}}
\newcommand{\pIm}{{\mathrm{Im}}}

\def\simleq{\; \raise0.3ex\hbox{$<$\kern-0.75em
      \raise-1.1ex\hbox{$\sim$}}\; }
\def\simgeq{\; \raise0.3ex\hbox{$>$\kern-0.75em
      \raise-1.1ex\hbox{$\sim$}}\; }

\begin{document}

\preprint{EHU-FT/000?}
\draft
\title{Color conductivity and ladder summation in hot QCD}
\author{J. M. Mart\'\i nez Resco \thanks{\tt wtbmarej@lg.ehu.es}, \\ 
        M. A. Valle Basagoiti \thanks{\tt wtpvabam@lg.ehu.es}}

\address{Departamento de F\'\i sica Te\'orica, \\ 
 Universidad del Pa\'\i s Vasco, Apartado 644, E-48080 Bilbao, Spain}

\date{\today}

\maketitle

\date{today}

\begin{abstract}

The color conductivity is computed at leading logarithmic order using a Kubo 
formula. We show how to sum an infinite series of planar ladder diagrams, 
assuming some approximations based on the dominance of 
soft scattering processes between hard particles in the plasma. The result 
agrees with the one obtained previously from a kinetical approach.    

\end{abstract}



\newpage
\section{Introduction}

Color conductivity has turned out to be a significant quantity for it enters 
directly into the theory governing the dynamics of long wavelength excitations 
in the QCD plasma. This was not obvious a priori, because a hydrodynamical 
description of the plasma usually involves transport coefficients associated 
with relaxation processes of momentum degrees of freedom for which the collision 
frequency is $\nu_{p} \sim g^{4} T \log(1/g)$~\cite{baym}, while the collision 
frequency for color relaxation is $\nu_{c} \sim g^{2} T \log(1/g)$~\cite{selikhov}. 
However, a major achievement in understanding the dynamics of very soft 
excitations in the QCD plasma, has been B\"odeker's finding of an effective 
theory described by a Langevin equation, in which thermal noise is parametrized 
by the static color conductivity~\cite{bodeker,bodeker3}.
Although he did not use explicitly the concept of color conductivity, the leading
logarithmic order result was implicit in his computation. As an intermediate step 
towards B\"{o}deker's theory, Arnold, Son and Yaffe~\cite{arnold} have proposed a 
Boltzmann equation from which color conductivity can be obtained. Along the way, 
some points of previous studies of color conductivity~\cite{selikhov,heiselberg}
have been clarified. Later, a rigorous derivation of the Boltzmann equation, starting 
from the quantum field equations, was given in Ref.~\cite{blaizot}. The Boltzmann 
equation leading to B\"{o}deker's theory (and color conductivity) can also be 
obtained starting from classical transport theory \cite{litim} or from the kinetic 
formulation of the hard thermal loop (HTL) effective theory \cite{valle}. Furthermore, 
a recent diagrammatic analysis by B\"odeker~\cite{bodeker1}  of the polarization 
tensor beyond the hard thermal loop approximation has also displayed the form of 
the Boltzmannn equation. Some calculations of the color conductivity at 
next-to-leading log order have appeared over the last year~\cite{blaizot1,arnold1}.

These studies of color conductivity at leading-log order have used a kinetic 
approach. Other transport coefficients in hot gauge theories have also been 
studied over the last years~\cite{baym,baym1,heiselberg1,heiselberg} using 
this approach. But there is an alternative way to compute transport coefficients, 
namely, through the use of Kubo formulas. However, this approach poses a difficulty: 
a summation of an infinite series of diagrams is required. Only for a scalar 
theory has this summation been carried out explicitly~\cite{jeon}. The higher 
order contributions were shown to come from ladder diagrams. In addition, the 
equivalence under certain conditions of a kinetic equation to the field theoretical 
computation was established~\cite{jeon,jeon1}. For the case of gauge theories it 
has been suggested that this kind of diagrams contributes at leading order~\cite{arnold}. 
However, because of the difficulty in carrying out these computations, this 
approach has not been pursued further. 

The purpose of this paper is to show how the ladder diagrams can be explicitly 
summed at leading-log order within the imaginary time formalism of thermal field 
theory. We will work at high temperature so the coupling constant is small and 
all zero temperature masses can be neglected.


\section{Color conductivity}

The static color conductivity associated with color flow may be defined by the 
constitutive relation $\langle j_{a}^i\rangle=\sigma_{a b}^{i j} E_{b}^j$, where 
$E_{b}^j$ are the components of an external, constant chromoelectric field and 
$\langle j_{a}^i\rangle$ is the ensemble average of the spatial part of the 
conserved color current density. 
Using linear response theory one may express color conductivity in terms of 
the low frequency, zero momentum limit of the retarded correlation function between 
color currents~\cite{mahan}. The corresponding formula is  
\be
 \sigma_{ij}^{ab}=-\frac{\partial}{\partial q^{0}}
 \pIm\,\Pi^{R \: ab}_{ij}(q^{0}, {\bf q}=0) \vert_{q^{0}=0}.
\ee
Because of isotropy, the dependence on the spatial and color indexes is very simple, 
$\sigma_{ij}^{ab}=\delta_{ij}\, \delta^{ab}\sigma_{c}$. The most efficient 
way to compute $\sigma_{c}$ from the above Kubo formula is to exploit 
the Lehmann representation, which provides a direct connection  via 
analytic continuation between 
the retarded Green's function and its counterpart in the imaginary 
time formalism, $\Pi^R(q^0,{\mathbf{0}})=\Pi(\nu_q=-i q^0+0^+,{\mathbf{0}})$.  
 
\begin{figure}[htb]
\centering
\leavevmode
\epsfysize=3cm
\epsfbox{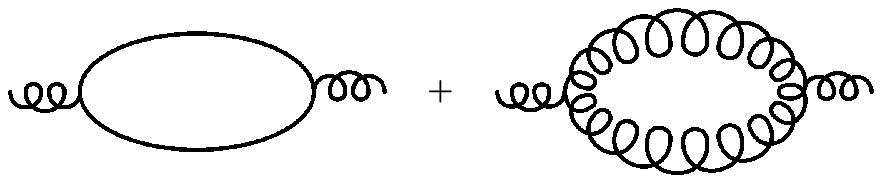}
\caption{One-loop contribution to gluon polarization tensor.}
\label{fglo}
\end{figure}

\begin{figure}
\centering
\leavevmode
\epsfysize=3cm
\epsfbox{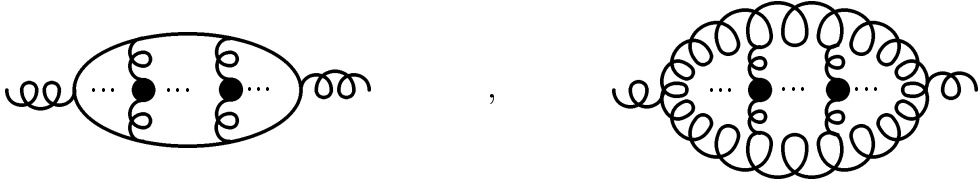}
\caption{Ladder diagrams contributing at leading order to gluon polarization tensor. The gluon 
line with the blob denotes the HTL resummed propagator.}
\label{fgladder}
\end{figure}

The dominant contribution would come from the diagrams in Fig.~\ref{fglo} with
all other diagrams of higher order in the loop expansion being subleading a
priori. However all ladder diagrams of Fig.~\ref{fgladder} will contribute at
the same order, as will be  shown in our explicit computation. 
This same kind of diagrams has been shown by 
Jeon~\cite {jeon} to give contributions of the same order in $\lambda$ as the 
one-loop graph for the viscosity in $\lambda \phi^4$ theory. In the context of 
gauge theories, it has been suggested~\cite{arnold} that the same kind of diagrams 
gives the leading order contribution. This can be understood in the following way.
In the limit of zero spatial momentum and small frequencies, 
there are going to be pairs of propagators which will carry the same 
momentum, and this would lead to an on-shell singularity for 
each product of propagators with nearly the same momenta. 
The solution is to include in the 
propagators of the side rails their thermal widths $\Gamma$. Thus, for each 
pair of propagators, the singularity  is replaced
with a term $1/\Gamma$. Hence, for a general ladder graph with $n$ rungs 
we will have
$g^2 (1/\Gamma)^{n+1} (g^2)^n$ where the first $g^2$ comes from the external 
vertices and each rung introduces a factor $g^2$ \cite{jeon,smilga}. 
For gauge theories, the  
thermal widths for hard momenta are proportional to $g^2 T \log(1/g)$, where 
the logarithm arises from the infrared behavior of the theory. As we shall 
see, a similar logarithm is produced by each integration over the 
small momentum transfer in the rung. So, both the 
one-loop and the ladder graphs are proportional to 
$g^2 T^2 q^0/\Gamma$.
Other ladder diagrams with two gluons lines in the rung are subleading because the
four gluon vertex has one more power of the coupling constant than the three
gluon vertex. 

Although we lack a rigorous power-counting scheme to dismiss all other 
possible higher order diagrams in the loop expansion, we do not expect them to
contribute at leading order, since we are not aware of any other way in which the 
coupling constants of each new vertex could be compensated. 
 
We now turn to the main point in this work, that is, how to sum the ladder diagrams 
of Fig.~\ref{fgladder}. First, we study the case when the side rails 
of the ladder are quark propagators.


\subsection{The quark contribution}

\begin{figure}
\centering
\leavevmode
\epsfysize=4cm
\epsfbox{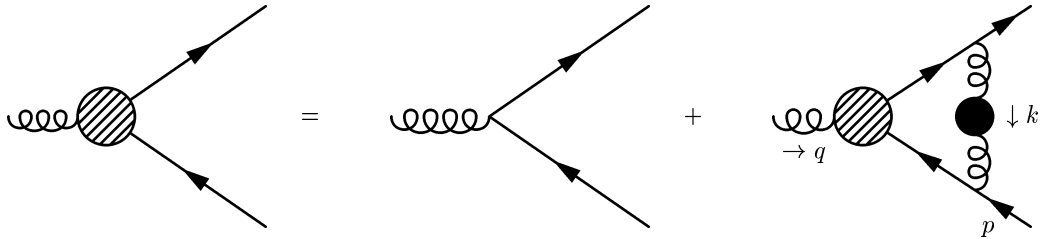}
\caption{Equation for the effective vertex which sums the quark ladder contributions.}
\label{fgvertquark}
\end{figure}

To include the thermal width in the propagators of the side rails, we modify the 
quark propagator by changing the delta functions of the spectral density for the free 
propagator by Lorentzians of width $2\gamma_{f}$
\be
\rho(\omega,\pv)=S_{+}(\pvuni) 
            \frac{2 \gamma_{f}}{(\omega-\modp)^{2}+\gamma_{f}^{2}} + 
	     S_{-}(\puni) 
            \frac{2 \gamma_{f}}{(\omega+\modp)^{2}+\gamma_{f}^{2}} \,,
\ee
where $S_{\pm}(\pvuni) =(\gamma^{0} \mp \gamma \cdot \pvuni)/2 $ and $\gamma_{f}$ 
is the fermion damping rate at hard momentum~\cite{pisarski}
\be \label{drf}
\gamma_{f}=\frac{N_{c}^{2}-1}{2N_{c}}\frac{g^{2}}{4 \pi} T \ln(1/g).  
\ee
From this spectral function the fermion propagator follows, as 
\bea
 S(\omega_{n},\pv) &=& \int_{-\infty}^{\infty}\frac{d\omega}{2\pi} 
 \frac{\rho(\omega,\pv)}{i\omega_{n}-\omega}= 
 S_{+}(\pvuni)\Delta_{+}(p)+S_{-}(\pvuni)\Delta_{-}(p)
 \nonumber \\
 &=&\frac{S_{+}(\pvuni)}{i \omega_{n}-\modp +i\gamma_{f}\sg{\omega_{n}}}
   +\frac{S_{-}(\pvuni)}{i \omega_{n}+\modp +i\gamma_{f}\sg{\omega_{n}}}.
\eea
To sum the ladder contributions we write the following equation for an effective vertex
\bea  \label{vquark}
 \Gamma^{i,a}(p+q,p)=\Gamma_{0}^{i,a}&+&g^{2}T\sum_{n}\int\frac{d^{3}\kv}{(2 \pi)^{3}} 
 \nonumber \\ 
 &\times& \gamma^{\alpha} t^{b}S(k+p+q)\Gamma^{i,a}(k+p+q,k+p)S(k+p)\gamma^{\beta}t^{b}
 D_{\alpha\beta}(k),
\eea
where the zero components of the momenta are $i$ times the corresponding 
Matsubara frequencies.  
$\Gamma_{0}^{i,a}=\gamma^{i} t^{a}$  is the tree level interaction vertex 
and $t^{a}$ are the $SU(N_{c})$ generators in the fundamental representation, 
with ${\mathrm{tr}}(t^{a}t^{b})=\delta^{a b}/2$. The gluon propagator 
$D_{\alpha \beta}$ is the HTL resummed propagator whose inclusion will 
be shortly justified. This equation is shown diagramatically 
in Fig.~\ref{fgvertquark}. 
To solve it exactly is beyond all hope, so we will try to extract the dominant
behaviour by making some approximations. 

The inclusion of the HTL propagator in the 
rung is based on the fact that the  interaction of two  
particles must be modified due to many body effects for small momentum 
transfer $k \ll T$. It has the form 
\be
   D_{\alpha\beta}(k)=\cP^{T}_{\alpha\beta}(\kvuni)\frac{1}
   {\omega_{n}^{2}+\modk^{2}+\Pi_{T}(k)}+
                       \delta_{4\alpha}\delta_{4\beta}\frac{1}{\modk^{2}+\Pi_{L}(k)},  
  \ee
where 
  $\cP^{T}_{ij}(\kvuni)=\delta_{ij}-\kuni_{i}\kuni_{j},
\ \cP^{T}_{4\alpha}=\cP^{T}_{\alpha4}=0$.
In order to 
perform the Matsubara sum in Eq.~(\ref{vquark}), it is 
convenient to introduce the spectral representation of this propagator
\begin{equation}
 D_{\alpha\beta}(\omega_{n},{\mathbf{k}})= 
 \frac{1}{{\mathbf{k}}^2} \delta_{4\alpha}\delta_{4\beta}+
 \int_{-\infty}^{\infty}\frac{d\omega}{2\pi} 
 \frac{1}{i\omega_{n}-\omega}\left[ -\rho_{T}(\omega,{\mathbf{k}} ) 
 {\mathcal{P}}_{\alpha\beta}^T + 
 \rho_{L}(\omega,{\mathbf{k}}) \delta_{4\alpha}\delta_{4\beta} 
 \right] ,
\end{equation}
and a  double spectral representation for the vertex,  whose general  
form parametrised by two functions 
$ F^{i a}_{1,2}(\omega_{1}, \omega_{2})$ is given by  
\begin{equation}
\Gamma^{i,a}(p+q,p)=
\int_{-\infty}^\infty \frac{d\omega_1}{2\pi}\frac{d\omega_2}{2\pi}  
 \left[\frac{F^{i a}_1(\omega_1,\omega_2)}{(p^0-\omega_1)(q^0-\omega_2)} + 
       \frac{F^{i a}_2(\omega_1,\omega_2)}{(p^0+q^0-\omega_1)(q^0-\omega_2)} \right],
\end{equation}  
Inserting the spectral representations into Eq.~(\ref{vquark}), 
this reduces in the high temperature limit $\omega \ll T$ to 
\bea  \label{vquark2}
 \Gamma^{i,a}(\omega_p+\nu_q,\omega_p; \pv)=
 \Gamma_{0}^{i,a}&-&g^{2}T \int \frac{d\omega}{2 
 \pi}
  \frac{d^{3}\kv}{(2 \pi)^{3}} \gamma^{\alpha} 
 t^{b}S(\omega_{p}+\nu_{q},{\mathbf{k}}+{\mathbf{p}})
 \nonumber \\ 
 &\times& \Gamma^{i,a}(\omega_p+\nu_q,\omega_p;{\mathbf{k}}+{\mathbf{p}})
 S(\omega_{p},{\mathbf{k}}+{\mathbf{p}})\gamma^{\beta}t^{b}
 \,\frac{\rho_{\alpha\beta}(\omega, {\mathbf{k}})}{\omega},
\eea
where the factor $T/\omega$ comes from the term containing the 
Bose-Einstein distribution $N(\omega)$ produced in the summation and 
$\omega_{p}, \nu_{q}$ denote the fermionic and bosonic Matsubara frequencies 
of the external lines. Thus, the summation has been traded for an integration over
$\omega$. 

To proceed further in the integration, let us recall 
the relevant kinematical domain 
from which arises the logarithmic sensitivity 
of the hard damping rates to the 
magnetic mass of order $g^2 T$. As we shall see, the leading-log order in 
the color conductivity results from the momentum transfer in the same domain. 

As is well known, the transverse  part of 
the interaction is dynamically screened due to 
Landau damping of the magnetic virtual gluons in the regime $|\omega|\leq 
|{\mathbf{k}}|$, and static Debye screening cuts off effectively the 
small angle divergences  of the longitudinal interaction. In terms of the 
gluon spectral density, the $(k,\omega)$ integral expressing the hard 
thermal damping 
rate~(\ref{drf}) is~\cite{pisarski,thoma,lifetime} given by 
\begin{equation}\label{int}
 \gamma_f=\frac{N_c^2 -1}{2 N_c} \frac{g^2 T}{4 \pi}
 \int_{\Lambda_{\mathrm{min}}}^{\Lambda_{\mathrm{max}}} dk\,k \int_{-k}^k 
 \frac{d\omega}{2 \pi \omega}\left[ \rho_{L}(\omega, k) +
       \right(1-\frac{\omega^2}{k^2}\left) \rho_{T}(\omega, k) \right] ,    
\end{equation}
where the lower cutoff $\Lambda_{\mathrm{min}}$, which is taken of the same order 
as the hypotetical magnetic mass $\sim g^2 T$, regularizes the 
infrared divergence of the integrand. The leading-log contribution 
to the damping rate arises from the exchange of quasistatic magnetic 
gluons in the $\omega \ll k$ limit. The reason for this is that 
the denominator of the integrand $k \rho_T/\omega$ vanishes in the 
static limit for $k=0$, while the denominator of $k \rho_L/\omega$ remains finite 
in this limit. In that domain, we can use the 
approximate expressions for the spectral densities
\begin{eqnarray}
    \rho_{T}(\omega,{\mathbf{k}})&=&\frac{8 \pi m_D^2\omega k}{16 k^6+ 
       \pi^2 m^4 \omega^2}, \\
    \rho_{L}(\omega,{\mathbf{k}})&=&\frac{\pi m_D^2\omega}{k (k^2+  
       m_D^2)^2} ,  
\end{eqnarray}
where 
$m_{D}^{2}=(N_{f}+2N_{c})g^{2}T^{2}/6$ is the Debye mass. 
Inserting them
into Eq.~(\ref{int}) produces a dominant logarithmic 
term 
\begin{equation}\label{cutoff}
 \gamma_f=\frac{N_c^2 -1}{2 N_c} \frac{g^2 T}{2 \pi^2} 
 \int_{\Lambda_{\mathrm{min}}}^{\Lambda_{\mathrm{max}}} \frac{dk}{k} 
 \arctan\left(\frac{\pi m_D^2}{4 k^2} \right) 
 \approx \frac{N_c^2 -1}{2 N_c} \frac{g^2 T}{4 \pi} 
 \int_{\Lambda_{\mathrm{min}}}^{\Lambda_{\mathrm{max}}} 
  \frac{d k}{k}
\end{equation}
coming enterely from the transverse part, together with 
a finite term coming from the longitudinal part which is 
of order $g^2 T$. Here, we only need the most singular logarithmic term, but 
for a computation of other transport coefficients, all 
contributions are possibly required. The upper cutoff 
$\Lambda_{\mathrm{max}}$ must be chosen of order $g T$ because 
when $k \gg g T$, the contribution from hard momentum transfer 
$k \sim T$ is computed using a non resummed gluon propagator. Thus
the contribution to the damping rate from this domain is 
proportional to 
\begin{equation}
        g^4 T^3 \int_{\Lambda_{\mathrm{max}}}^\infty \frac{dk}{k^3}.
\end{equation}
If the upper cutoff were chosen of order $T$ in Eq.~(\ref{cutoff}), 
one would make the mistake of relying on an integrand which in the 
range $g T \leq k \leq T$ does not contain the correct dependence on 
the momentum transfer. So the result would be wrong by a factor of two. 
(The correct spectral density in that range is proportional to 
$m_D^2 \omega/(k^3 (k^2-\omega^2))$.)

Briefly, concerning the $(k,\omega)$ integration 
in the vertex equation~(\ref{vquark2}), this can be computed by retaining 
only the transverse part, giving the result
\begin{equation}
           \int_{-k}^k \frac{d \omega}{2 \pi} \frac{\rho_T}{\omega} 
         \approx 
        \frac{1}{k^2}.    
\end{equation} 
This result is the same we would have obtained if, in the 
initial vertex equation~(\ref{vquark}), we had retained only 
the static mode in the sum with the substitution
$D_{\alpha \beta} \rightarrow (\delta_{i j} - \hat{k}_i \hat{k}_j) 
1/k^2$.  
The following integration in $k$ must be cut off by a 
semihard scale $\Lambda_{\mathrm{max}} \sim g T$ on one side and 
a soft scale $\Lambda_{\mathrm{min}} \sim g^2 T$ on the other side. 
In addition, for a pair of fermionic external lines nearly on shell, 
for example $p^0 \approx |{\mathbf{p}}|, p^0+q^0 \approx |{\mathbf{p}}|$, 
we have only to retain in the vertex equation~(\ref{vquark2})
one of the nearly singular products, $\Delta_+ \Delta_+$ in that 
case.  Of course, there is another pair of values of the external lines 
leading to a nearly singular piece, 
$p^0 \approx -|{\mathbf{p}}|, p^0+q^0 \approx -|{\mathbf{p}}|$, for 
which only the product $\Delta_- \Delta_-$ must be retained. This  
additional simplification is due to the fact that when $k \ll p$, 
a small  momentum transfer cannot change the mass shell of the 
hard particles involved in the scattering.   

Now, we make the following ansatz for the efective vertex \footnote{The posible
structure of the vertex is greatly simplified by the fact that $\qv=0$.}
\bea
 \Gamma^{i,a}(p+q,q)=(A \gamma^{i} +B \gamma^{0} \puni^{i} 
 +C \gamma^{j} \puni^{i} \puni^{j} ) t^{a},
\eea
where $A,B,C$ are some scalar functions which can depend on 
$p^{0}, q^{0}$ and $\modp$. 
First we consider the term with $\Delta_{+}\Delta_{+}$.
The equation for the efective vertex is now
\bea
 \Gamma^{i,a}(p+q,q)&=&
 \Gamma_{0}^{i,a}-\frac{g^{2}T t^{a} }{2N_{c}} \int \!\!\frac{d^{3}\kv}{(2 \pi)^{3}}
 \: \Delta_{+}(p+k+q)\Delta_{+}(p+k) 
 \nonumber  \\ 
 &&\times(A+B+C)(\gamma^{0}-\gv \cdot \kvuni (\pkvuni) \cdot \kvuni) (\pkuni)^{i} 
 \frac{1}{\modk^2}.
\eea
The next step is to compute the integrals over the momentum transfer $\kv$. It is useful 
to introduce a momentum variable $u$, defined as $u^2=(\kv+\pv)^2$, so that 
\be
 \int \frac{d^3 \kv}{(2 \pi)^3}\ldots \rightarrow 
 \frac{1}{(2\pi)^2 p}\int_{\Lambda_{\mathrm{min}}}^{\Lambda_{\mathrm{max}}} dk\, k 
 \int_{|k-p| }^{k+p} du \, u \ldots
\ee
One may decouple the limits of integration in the following way. In the limit 
${\mathbf{q}}=0,q^0 \rightarrow 0$, with the pair of external fermion lines nearly 
on-shell at hard momentum, $p^0 \sim \modp \sim T$, we are concerned 
with a pair of internal fermionic lines sharing nearly the same loop momenta, 
so we can anticipate that values of $u\sim p^0 $ and $u\sim p^0+q^0$ are important 
in the integration process. Since most of the contribution is in the region 
$u \sim p^0$, the limits on the integral over $u$ can be extended to $\pm \infty$, 
with the contribution outside the initial range being small. 
Furthermore, if the scalar functions $A,B,C$ are not strongly dependent on $p$ we can 
take them out of the integral, along with the factor $u$, evaluated in $u=\modp$. 
Given these approximations, the $u$ integration results in  
\be   \label{colineal}
 \int_{-\infty}^{\infty} du \frac{1}{i\omega_{p}-u+i\gamma_f \sg{\omega_p}}\,
 \frac{1}{i\omega_{p}+i\nu_{q}-u+i\gamma_f \sg{\omega_p+\nu_q}}= 
 \frac{2\pi \cD(\omega_p,\nu_q)}{|\nu_q|+2\gamma_f} ,
\ee
where $\cD(\omega_p,\nu_q)=\theta(-\omega_p)\theta(\omega_p+\nu_q)+
\theta(\omega_p)\theta(-\omega_p-\nu_q )$. We see that the thermal width regulates a
singular behaviour when $\nu_{q}$ vanishes. Upon completion of all these integrations, 
it is straightforward to solve for 
the scalar functions. Finally, we obtain for the vertex function when the external 
fermions are nearly on shell at leading logarithmic order 
\be
 \Gamma^{i a}(\omega_p+\nu_q,\omega_p; \pv)=\gamma^{i}t^{a}- 
 \frac{2\gamma_{f}\cD(\omega_p,\nu_q)}
 {(N_{c}^{2}-1)|\nu_q|+2N_{c}^{2}\gamma_{f}}\gamma^{0}\puni^{i}t^{a} , 
\ee
where the logarithmic dependence on $g$ coming from the $k$ integral has been 
replaced by the same logarithm from the fermion damping rate. This result is 
consistent with the assumption about the smooth dependence on $\modp$ of the 
scalar functions of the vertex. If we consider the term $\Delta_{-}\Delta_{-}$ 
we arrive at the same result. Here it should be emphasized that the vertex correction 
arising from ladder summation is of the same order as the one at tree level. 

It remains to compute the polarization tensor
\be
 \Pi^{i j}_{a b}(\nu_q,{\mathbf{0}})=g^2 N_f T \sum_{\omega_p}\int 
 \frac{d^3{\mathbf{p}}}{(2 \pi)^3} {\mathrm{tr}} \left[
 \Gamma^{i\,a}(\omega_p+\nu_q,\omega_p; {\mathbf{p}}) 
 S(\omega_p,{\mathbf{p}}) \gamma^j t^b 
 S(\omega_p+\nu_q,{\mathbf{p}}) \right] . 
\ee
The most difficult task is to do the Matsubara sum. It will be useful to introduce 
a double spectral representation of the product of Green's functions as follows. 
Let $\Delta(p+q,p) = G(p+q) G(p)$, where the Green's function $G(p)$ has a single 
spectral representation 
\be
 G(p)=\int_{-\infty}^\infty \frac{d\omega}{2 \pi} 
           \frac{\rho(\omega,\pv)}{p^0-\omega} .
\ee
Then, it is easy to check that $\Delta(p+q,p)$ admits the double spectral representation 
\be\label{doble}
 \Delta(p+q,p) = \int_{-\infty}^\infty \frac{d\omega_1}{2\pi}\frac{d\omega_2}{2\pi}  
 \left[\frac{F_1(\omega_1,\omega_2)}{(p^0-\omega_1)(q^0-\omega_2)} + 
       \frac{F_2(\omega_1,\omega_2)}{(p^0+q^0-\omega_1)(q^0-\omega_2)} \right],
\ee
where the corresponding $F$ functions are
\bea  \label{dospecone1}
 F_{1}(\omega_{1},\omega_{2})&=&
 \rho(\omega_{1}+\omega_{2}; \pv+\qv)\rho(\omega_{1}; \pv),  
 \\  \label{dospecone2}
 F_{2}(\omega_{1},\omega_{2})&=&
 -\rho(\omega_{1}; \pv+\qv)\rho(\omega_{1}-\omega_{2}; {\mathbf p}).
\eea
Similarly, the product of the vertex correction and two propagators,
\be
 \delta \Delta(p+q,p)=\frac{{\mathcal D}(\omega_p,\nu_q)}{b+c \vert \nu_q \vert } \;
 \frac{1}{ i \omega_p - \modp +
          i \gamma\, {\mathrm sgn}(\omega_p) } \;
 \frac{1}{ i \omega_p + i \nu_q - |{\mathbf p}| +
        i \gamma\, {\mathrm sgn}(\omega_p+\nu_q) },
\ee 
has the same representation with the $\delta F$ functions  
\begin{eqnarray}   \label{dospecres1}
\delta F_{1}(\omega_{1},\omega_{2})&=&
\frac{2b (\gamma^{2}+|{\mathbf p}|^{2}+\omega_{1}^{2}-
          2 |{\mathbf p}|\, \omega_{1}- |{\mathbf p}|\, \omega_{2}+
           \omega_{1} \omega_{2})-c \gamma \omega_{2}^{2}}
     {(\gamma^{2}+(\omega_{1}-|{\mathbf p}|)^{2})
      (\gamma^{2}+(\omega_{1}-|{\mathbf p}|)^{2}
       -2 |{\mathbf p}|\, \omega_{2}+2\omega_{1}\omega_{2}+
       \omega_{2}^{2})(b^{2}+c^{2}\omega_{2}^{2})},\\  
 \label{dospecres2}
\delta F_{2}(\omega_{1},\omega_{2})&=&
\frac{-2b (\gamma^{2}+|{\mathbf p}|^{2}+\omega_{1}^{2}-
          2 |{\mathbf p}|\, \omega_{1}+ |{\mathbf p}|\, \omega_{2}-
           \omega_{1} \omega_{2})-c \gamma \omega_{2}^{2}}
     {(\gamma^{2}+(\omega_{1}-|{\mathbf p}|)^{2})
      (\gamma^{2}+(\omega_{1}-|{\mathbf p}|)^{2}
       +2 |{\mathbf p}|\, \omega_{2}-2\omega_{1}\omega_{2}+
       \omega_{2}^{2})(b^{2}+c^{2}\omega_{2}^{2})}.
\end{eqnarray}

At this point, we have all the elements required. There are two dominant contributions 
coming from the two products of internal propagators whose poles are nearly coincident, 
so in the following, we write twice the contribution corresponding to one of them 
(in particular, the one that has the singularity in $p^0 = |{\mathbf p}|$, see 
Eqs.~(\ref{F2}) and~(\ref{deltaF2}) below). Making use of the double spectral representation 
to do the Matsubara sum and the fact that the angular dependence of the integrand 
is trivial, one obtains
\begin{eqnarray}
 \Pi^{i j}_{a b}(\nu_q,{\mathbf{0}})=& &\frac{2}{3} g^2 N_f \delta^{i j}
     \delta_{a b}
  \int \frac{d^3{\mathbf{p}}}{(2 \pi)^3} 
     \int_{-\infty}^\infty \frac{d\omega_2}{2\pi} 
     \frac{1}{i \nu_q - \omega_2}
       \int_{-\infty}^\infty \frac{d\omega_1}{2\pi} n_f(\omega_1) 
      \nonumber \\
  &&\times\left[F_1(\omega_1,\omega_2) +  F_2(\omega_1,\omega_2)+
     \delta F_1(\omega_1,\omega_2) + 
 \delta F_2(\omega_1,\omega_2) \right].
\end{eqnarray}
Now, it easy to make the analytic continuation $i\nu_q \rightarrow q^0+i0^+$ and 
to expand the imaginary part to lowest order in $q^0$. The result is 
\begin{equation}
 {\mathrm Im}\, \Pi^{R\, i j}_{a b}(q^0,{\mathbf{0}}) = 
    -\frac{1}{3} g^2 N_f q^0 \delta^{i j}\delta_{a b}
      \int \frac{d^3{\mathbf{p}}}{(2 \pi)^3} 
      \int_{-\infty}^\infty \frac{d\omega}{2\pi} n_f'(\omega) 
 \left[F_2(\omega,0)+\delta F_2(\omega,0)\right],  
\end{equation}
where we have used the property 
$(\delta)F_{1}(\omega_{1},\omega_{2})+(\delta)F_{2}(\omega_{1}+\omega_2,\omega_{2})=0$.
Thus, the explicit expressions of $F_{2}(\omega,0)$ and $\delta F_{2}(\omega,0)$ 
are required, 
\bea   \label{F2}
 F_2(\omega,0) &=& -\rho_+(\omega,{\mathbf p})^2 = 
 -\frac{4 \gamma_f^2}
  {\left((\omega-|{\mathbf p}|)^2+\gamma_f^2 \right)^2}\,, \\ 
 \label{deltaF2}
 \delta F_2(\omega,0) &=&  
 \frac{2}
 {N_c^2 \left((\omega-|{\mathbf p}|)^2+\gamma_f^2 \right)}\, ,
\eea
which, in the limit $\gamma_f \rightarrow 0$, can be replaced by 
$F_2(\omega,0)=-2\pi\delta(\omega-\modp)/\gamma_f$ and 
$\delta F_2(\omega,0)=2\pi\delta(\omega-\modp)/(N_c^2 \gamma_f)$. These substitutions 
give the final result to leading logarithmic order
\be
 {\mathrm Im}\, \Pi^{R\, i j}_{a b}(q^0,{\mathbf{0}})=
  -\frac{g^2 T^2 N_f}{36 \gamma_f}\, q^0 \delta^{i j}
     \delta_{a b} \left(1-\frac{1}{N_c^2}\right)= 
  -q^0 \delta^{i j} \delta_{a b}\frac{2 \pi}{9} \frac{N_f}{N_c} 
     \frac{T}{\log (1/g)} . 
\ee


\subsection{The gluonic contribution}

\begin{figure}
\centering
\leavevmode
\epsfysize=4cm
\epsfbox{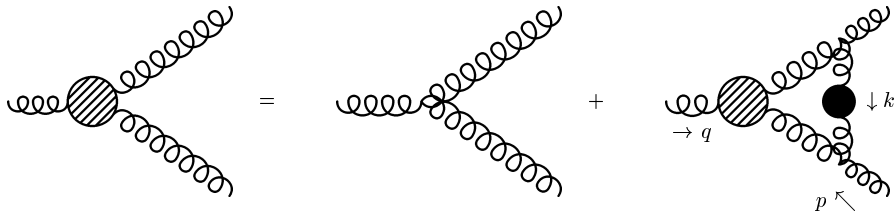}
\caption{Equation for the effective vertex which sums the gluon ladder contributions.}
\label{fgvertgluon}
\end{figure}

We only need to consider transverse gluons propagating 
in the side rails of the ladder, since the longitudinal ones 
do not propagate at hard momentum. 
We again modify the gluon propagator in order to include the effects of a thermal 
width $2\gamma_{g}$
\be\label{DD}
 \rho(\omega,\pv)=
 \frac{1}{\modp}\left(\frac{\gamma_{g}}{(\omega-\modp)^{2}+\gamma_{g}^{2}}+
 \frac{\gamma_{g}}{(\omega+\modp)^{2}+\gamma_{g}^{2}} \right),
\ee
where we use the gluon damping rate at hard momentum 
$\gamma_{g}=\alpha_{s}\, N_{c} T \, \ln(1/g)$. 
The vertex equation which sums the ladders is shown diagramatically in 
Fig.\ref{fgvertgluon}.
The normal vertex is 
\be
 \Gamma^{ijk}_{0\: abc}(q,p,-p-q)=if^{abc}[ 2 \delta^{jk} p^{i}-\delta^{ij} p^{k}-
 \delta^{ik} p^{j}].
\ee
We parametrize the effective vertex as
\be
\Gamma^{ijk}_{abc}(q,p,-p-q)=if^{abc}[ A p^{i} \cP^{T}_{jk}(\pvuni)+B \delta^{ij}p^{k}+
C\delta^{ik} p^{j}+D\frac{p^{i} p^{j} p^{k}}{\pv^{2}}].
\ee
We need the effective vertex only when the indices $j,k$ are contracted 
with the transverse projectors 
\be
 \Lambda^{ijk}_{abc}(q,p,-p-q)=\cP^{T}_{jn}(\pvuni)\Gamma^{inm}_{abc}(q,p,-p-q) 
 \cP^{T}_{mk}(\pvuni)=
 if^{abc}  p^{i} \cP^{T}_{jk}(\pvuni) A. 
\ee
After a bit of algebra the equation for the effective vertex reduces to
\be
 p^{i} A=2 p^{i}+2 g^{2}N_{c} \pv^{2}T \sum_{n} \int \frac{d^{3}\kv}{(2 \pi)^{3}}
 A (p+k)^{i} D^{t}(p+k+q) D^{t}(p+k) \frac{1}{\modk^2}, 
\ee
where $D^t(p)$ is the propagator obtained from the spectral density 
of Eq.~(\ref{DD}). Within the same previous 
approximations, one obtains 
\begin{equation}
  \Lambda^{i j k}_{a b c}(\omega_p+\nu_q,\omega_p;{\mathbf p}) = 
  2 i f^{a b c} p^i {\mathcal P}_{j k}^T  
  \left(1 + \frac{ \gamma_g {\mathcal D}(\omega_p,\nu_q)}
     {|\nu_q|+\gamma_g} \right) .
\end{equation}
With this result, following the same steps as earlier, it is easy to 
compute the imaginary part of the gluon polarization tensor. It reads
\begin{equation}
 {\mathrm Im}\, \Pi^{R\, i j}_{a b}(q^0,{\mathbf{0}})=
  -\frac{g^2 T^2 N_c}{9 \gamma_g}\, q^0 \delta^{i j} \delta_{a b} . 
\end{equation}

Finally, the sum of both contributions gives the correct value for the 
color conductivity $\sigma_c = \omega_{\mathrm p}^2/\gamma_g$, where
$\omega_{\mathrm p}^2\equiv \left(g^2 T^2/18\right) (2 N_c+N_f)$ is 
the plasma frequency.


\section{Summary}

In this work we have derived the color conductivity to the leading logarithmic 
order. The main point here has been the evaluation of the Kubo formula for the 
current-current correlator within the framework of thermal field theory in the 
imaginary time formalism. We have shown how to do the summation of the ladder 
vertex corrections, making use of some approximations based on the kinematical 
regime of the scattering, which corresponds to exchange of quasistatic transverse 
gluons of soft momentum, $ g^2 T< |{\mathbf k}| < g T$. We have also shown how
to introduce a double spectral representation of three-point functions in order 
to do the Matsubara sums involved in this formalism. 

The possibility of computing other transport coefficients such as viscosity or 
electrical conductivity by a similar procedure deserves further consideration. 
This kind of transport properties follows from an integral over the differential 
cross section multiplied by the square of the momentum transfer $k^2$. The 
corrections due to Debye screening and Landau damping arising at order $g T$ 
are sufficient to render this integral infrared finite, scaling as 
$g^4 \log(k^\ast/m_{{\mathrm el}}) \sim g^4 \log(1/g)$, where $ k^\ast$ is a 
scale separating semihard and hard momentum transfers, restricted by 
$g T \ll k^\ast \ll T$ but otherwise arbitrary. So the $\log(1/g)$ comes from 
the sensivity to the momentum scale $m_{\mathrm el} \sim g T$, while the $\log(1/g)$ 
in the color conductivity and the damping rate comes from a sensitivity to the 
magnetic scale $g^2 T$. This means that in order to attempt a similar computation 
of these transport coefficients, one must include into the rungs of the ladder 
the longitudinal part of interaction as well as the Landau damping effects. The 
infrared sensitivity to the scale $g^2 T$ should be compensated by a similar one 
coming from the hard damping rate.


\section*{Acknowledgments}
This work is partially supported by UPV/EHU grant 063.310-EB187/98 and 
CICYT AEN 99-0315. 
J. M. Mart\'\i nez Resco is supported by a Basque Government grant.



\newpage


\end{document}